\documentclass[a4paper]{jpconf}
\usepackage{graphicx}
\usepackage{amsmath}
\usepackage{latexsym}
\usepackage{amssymb}
\usepackage{tabularx}
\usepackage{subfigure}

\begin{document}

\title{Bar-mode instability suppression in magnetized relativistic stars}

\author{L. Franci$^{1}$, R. De Pietri$^{1}$, K. Dionysopoulou$^{2}$ and L. Rezzolla$^{2,3}$}
\address{$^{1}$ Dipartimento di Fisica e Scienze della Terra, Universit\`a di Parma and INFN, Parma, Italy}
\address{$^{2}$ Max-Planck-Institut f\"ur Gravitationsphysik, Albert-Einstein-Institut, Golm, Germany}
\address{$^{3}$ Institut f\"ur Theoretische Physik, Frankfurt am Main, Germany}
\ead{luca.franci@fis.unipr.it}

\begin{abstract}
We show that magnetic fields stronger than about $10^{15}$ G are able to
suppress the development of the hydrodynamical bar-mode instability in
relativistic stars. The suppression is due to a change in the rest-mass
density and angular velocity profiles due to the formation and to the
linear growth of a toroidal component that rapidly overcomes the original
poloidal one, leading to an amplification of the total magnetic
energy. The study is carried out performing three-dimensional
ideal-magnetohydrodynamics simulations in full general relativity,
superimposing to the initial (matter) equilibrium configurations a purely
poloidal magnetic field in the range $10^{14}-10^{16}$ G. When the seed
field is a few parts in $10^{15}$ G or above, all the evolved models show
the formation of a low-density envelope surrounding the star. For much
weaker fields, no effect on the matter evolution is observed, while
magnetic fields which are just below the suppression threshold 
are observed to slow down the growth-rate of the instability.
\end{abstract}

\section{Introduction}
Differentially rotating neutron stars (NSs) are subject to the so-called $m
= 2$ dynamical bar-mode instability for non-radial axial modes with
azimuthal dependence $e^{im\phi}$ ($m = 1$, $2$, ...), when the
instability parameter $\beta \equiv T/ |W|$ (i.e. the ratio between
the rotational kinetic energy $T$ and the gravitational binding energy
$W$) exceeds a critical value.
The bar-mode instability in differentially rotating magnetized NSs
 has already been studied in the Newtonian case by Camarda et al.
\cite{Camarda2009}, suggesting that the effect of magnetic fields on
the emergence of the instability is not likely to be very signiﬁcant
unless when NSs are born very highly magnetized.

In this work, we investigate if the presence of magnetic fields can
affect the onset and development of this kind of instability in full
general relativity, as well as the role played by the magnetic braking
to possibly suppress the instability.

Our study is motivated by the potential implications that strong
magnetic fields may have for astrophysical scenarios and for
gravitational wave astronomy. In particular, during the last decade a
general consensus has formed that long Gamma-Ray Bursts (GRBs) arise
from the collapse of massive stars while short GRBs, most likely, from
neutron star mergers. A key ingredient for both these categories of
astrophysical phenomena is the formation of magnetic structures that
could power the high-speed particle jets associated with GRBs
\cite{Rezzolla2011}.

We choose to evolve different equilibrium relativistic stellar
models that have already been studied in the non-magnetized case by
Baiotti et al.~\cite{Baiotti2007}, so that their behavior against the
bar-mode instability is already known and fully understood when no
magnetic fields are present. 

\begin{figure*}[1.0\textwidth,t]
  \begin{center}
    \includegraphics[width=0.32\textwidth]{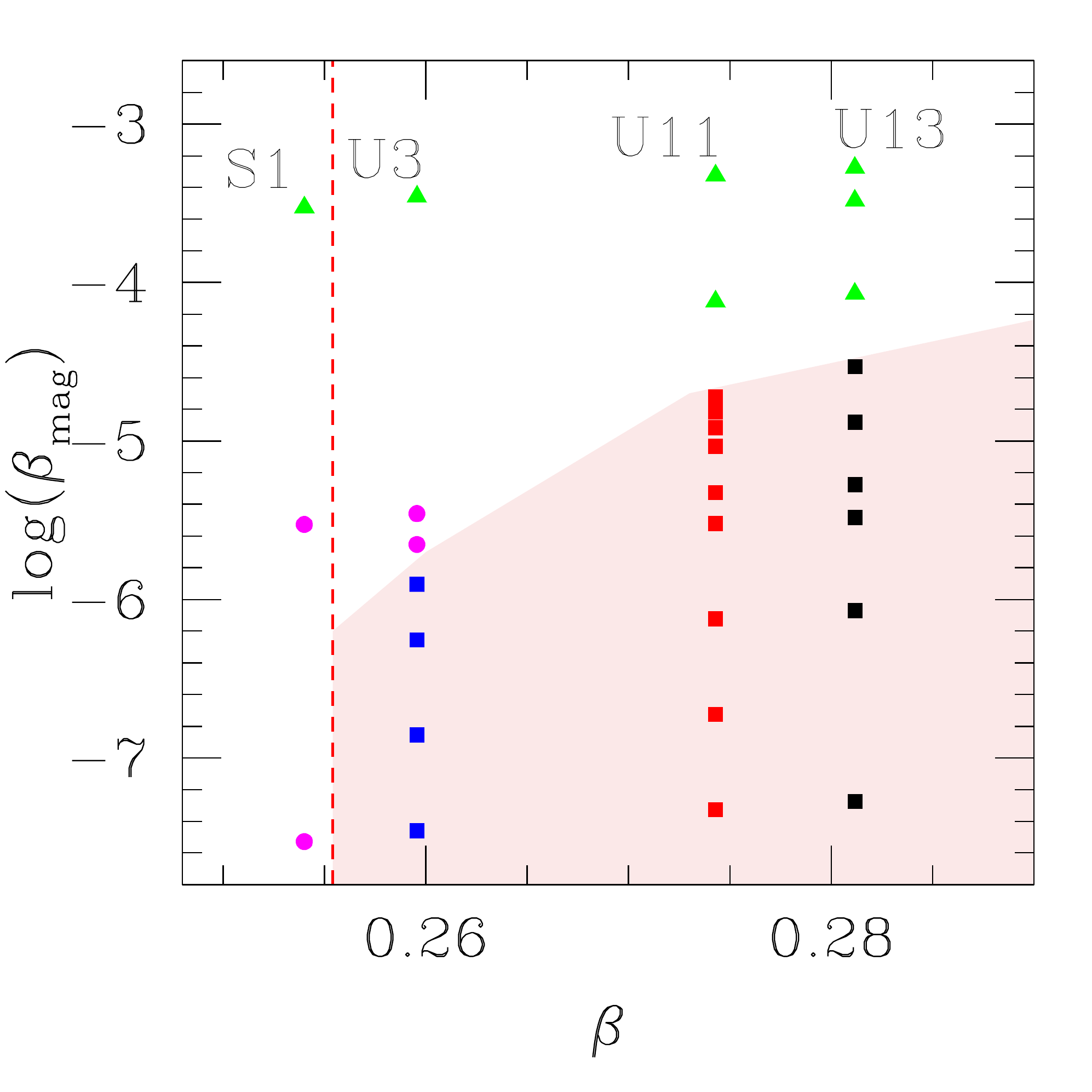}
    \includegraphics[width=0.32\textwidth]{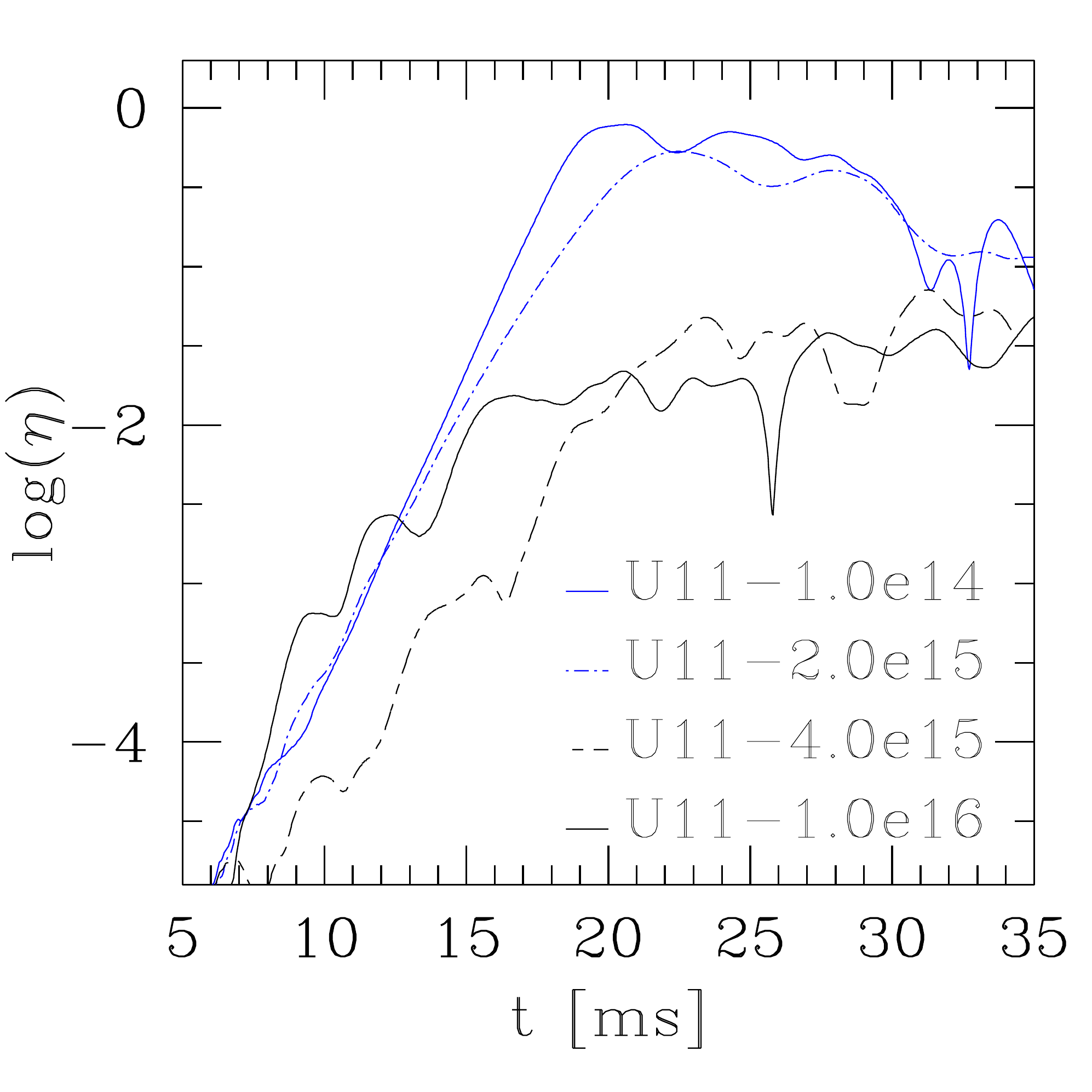}
    \includegraphics[width=0.32\textwidth]{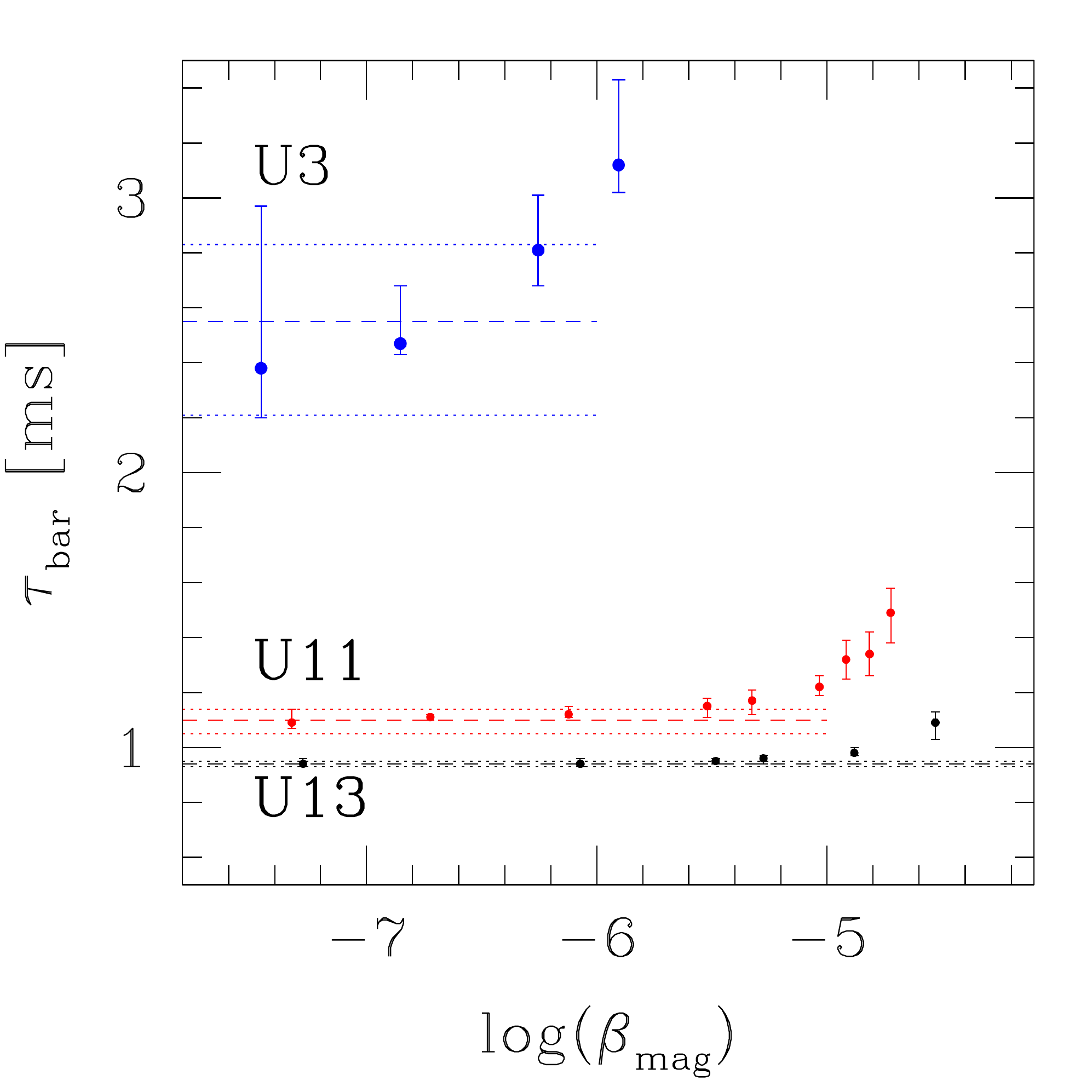}
  \end{center}
  \vglue-0.8cm
    \caption{(Left panel): parameter space
      $(\beta,\;\beta_{\text{mag}})$ with the initial values of these
      two quantities for all the simulated models: the models inside
      the red shaded region turned out to be still matter-unstable
      even when a magnetic field is present. 
      (Central panel): time evolution of the distortion parameter $\eta$ for
      model \texttt{U11} for four different values of the seed
      poloidal magnetic field strength. We use blue lines for
      matter-unstable models and black lines for the stable ones.
      (Right panel): growth-time $\tau_{bar}$ of the bar-mode instability versus the
      initial values of $\beta_{\text{mag}}$. The dashed
      lines represent the values in the non-magnetized cases while the
      dotted lines stand as error bars.}
    \label{fig:InitialModels}
\end{figure*}

\section{Initial data}
The initial data for our simulations are computed as stationary
equilibrium solutions of axisymmetric relativistic stars that are
rapidly differentially rotating. In particular,
we assume the non-uniform $j$-law angular velocity distribution 
\begin{equation}
\Omega_{c} -\Omega   =   \frac{1}{\hat{A}^{2} R_{e}^{2} }
       \left[ \frac{(\Omega-\omega) r^2   \sin^2\theta e^{-2\nu}
              }{1-(\Omega-\omega)^{2} r^2 \sin^2\theta e^{-2\nu}
       }\right] \,,
\label{eq:velocityProfile}
\end{equation}
where $R_{e}$ is the coordinate equatorial stellar radius, the
coefficient $\hat{A}$ is a measure of the degree of differential
rotation, while $\omega$ and $\nu$ are metric components in spherical 
coordinates. 

A seed poloidal magnetic field is 
added to these initial equilibrium models as a perturbation by
introducing a purely toroidal vector potential $A_\phi = A_b \left(
\max(p - p_\text{cut},0) \right)^2$, where $p_\text{cut}$ is $4 \%$ of
the maximum initial pressure $(p_\text{cut}=0.04 \;\max(p(0)))$ and
$A_b$ is chosen in such a way so that it corresponds to the maximum of
the initial magnetic field on the $(x,\,y)$ plane
$B^{z}_\text{max}|_{t,z=0}$. 

The equilibrium models considered here have been calculated using a
relativistic polytropic equation of state $p=K\rho^\Gamma$ with
$K=100$ and $\Gamma=2$, in analogy with previous works in the
literature. In particular, to better find out the changes to the
bar-mode dynamics induced by the presence of a magnetic field, we
focused our attention on a sequence of models having a fixed constant
amount of differential rotation $\hat{A}=1$ and a constant rest mass
$M_0\simeq 1.5\ M_{\odot}$, which were already studied in the
non-magnetized case in~\cite{Baiotti2007}. All the features of these
models are then described in further detail in Tab. I of
\cite{Baiotti2007} and here we use the notation therein reported. For
the present study, we selected the three models which were better
discussed therein, namely models \texttt{U3}, \texttt{U11} and
\texttt{U13}, where \texttt{U3} is the closest to the threshold for
the onset of the bar-mode instability and \texttt{U13} is almost the
fastest rotating equilibrium model which can be obtained for this
sequence at constant barionic mass (see the left panel of
Fig.~\ref{fig:InitialModels}, where the threshold is indicated with a
red dotted line corresponding to $\beta = 0.255$).

Indeed, all the simulated initial models are represented in the left
panel of Fig.~\ref{fig:InitialModels}, where they are identified in
terms of the initial values of the two parameters $(\beta
,\;\beta_{\text{mag}})$, defined as $\beta \equiv T/|W|$ and
$\beta_{\text{mag}}\equiv E_{\text{mag}}/(T+|W|)$, respectively. Here,
$T$ is the rotational kinetic energy, $W$ the gravitational binding
energy and $E_{\text{mag}}$ the total magnetic energy. Hereafter, we
will refer to a particular evolved stellar model by using the notation
U$xx$-$yy$, where $xx$ is the initial model and $yy$ denotes the
maximum initial magnetic field strength (\texttt{U11-1.0e14} refers to
a \texttt{U11} model with a super-imposed purely poloidal magnetic
field such that $B^{z}_\text{max}|_{t,z=0}=1.0 \times 10^{14}$ G).

\begin{figure*}[t!]
\begin{center}
\vbox{\hbox{
\includegraphics[width=0.24\textwidth]{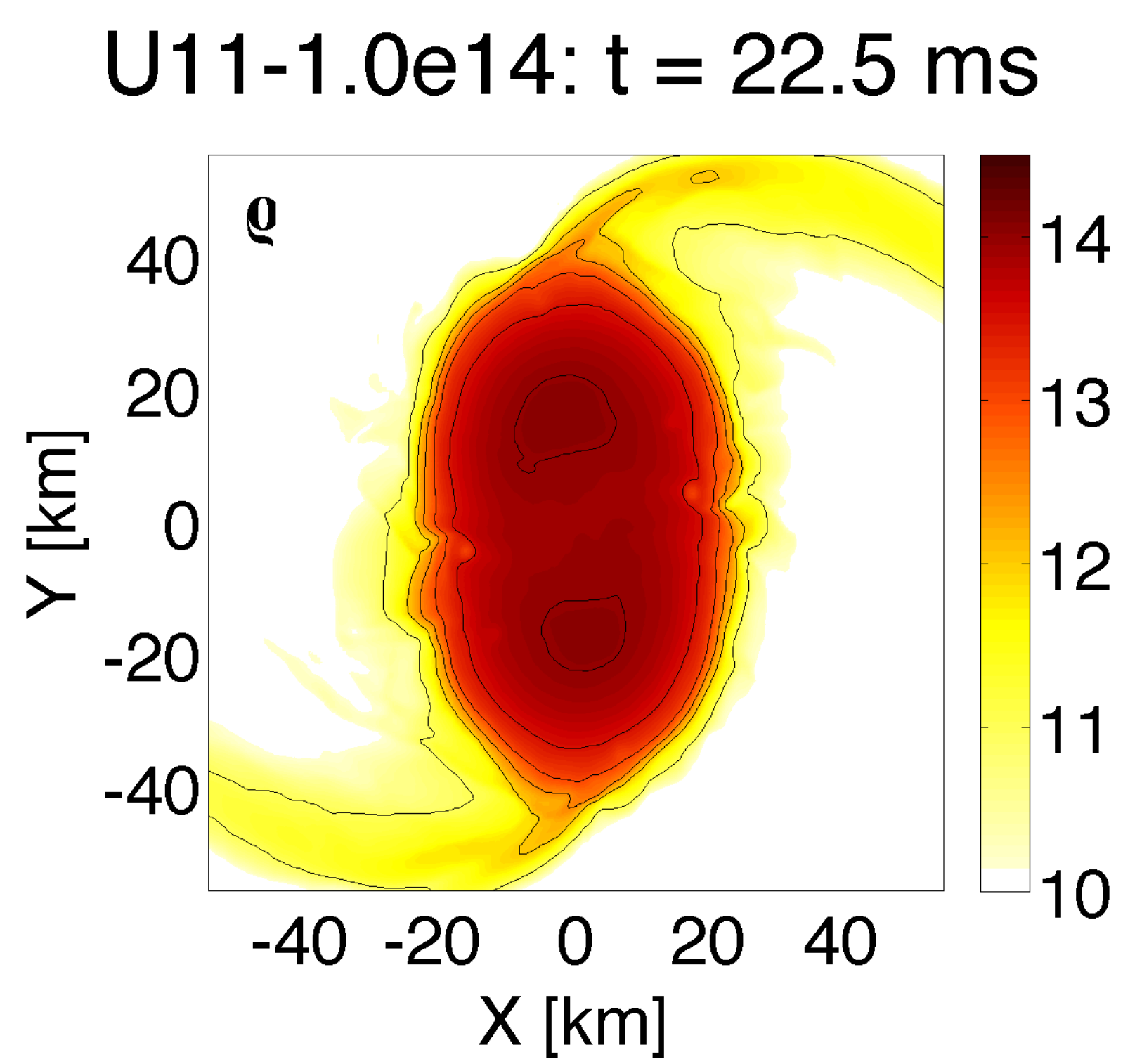}
\hspace{-0.1cm}
\includegraphics[width=0.24\textwidth]{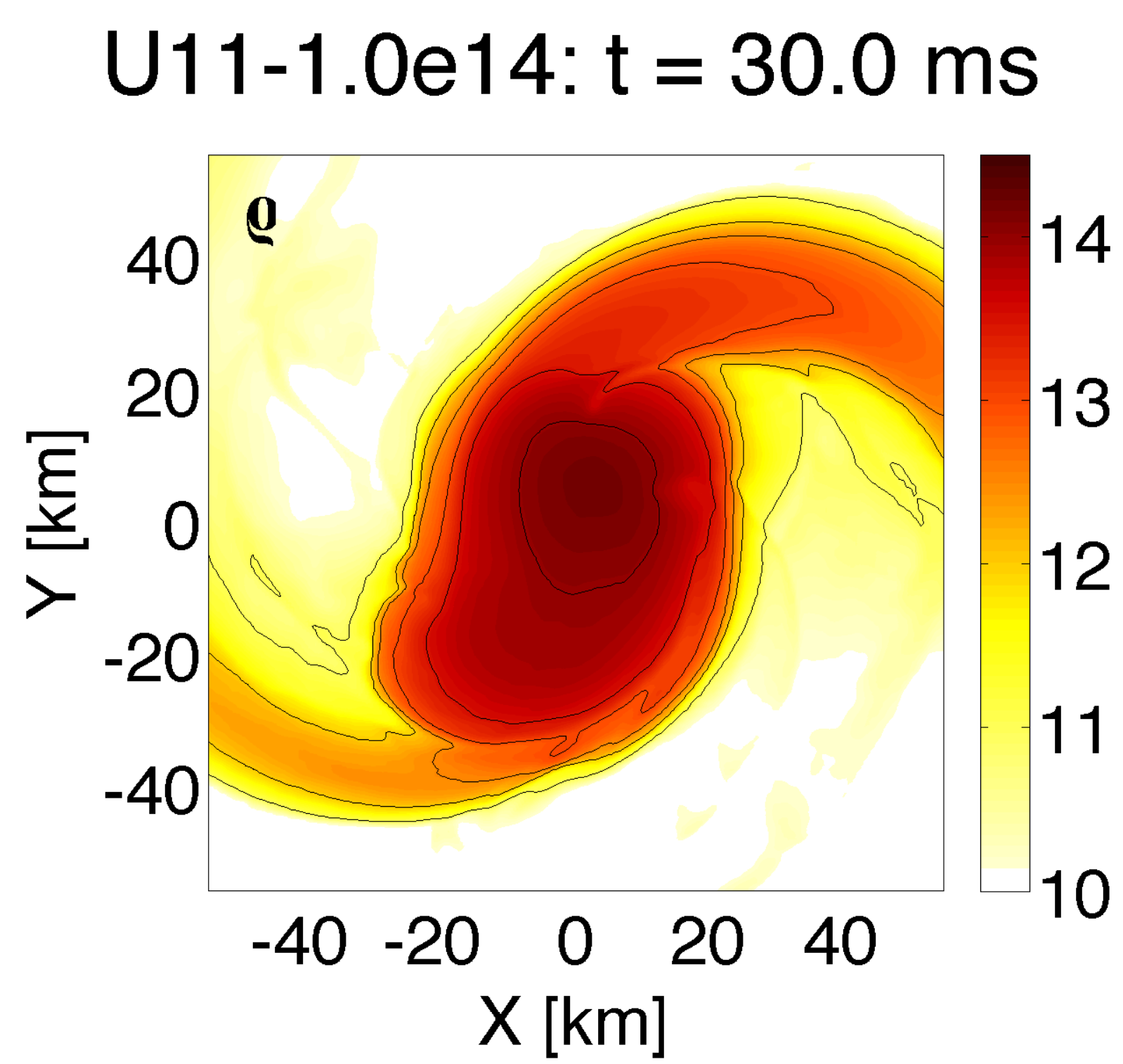}
\includegraphics[width=0.24\textwidth]{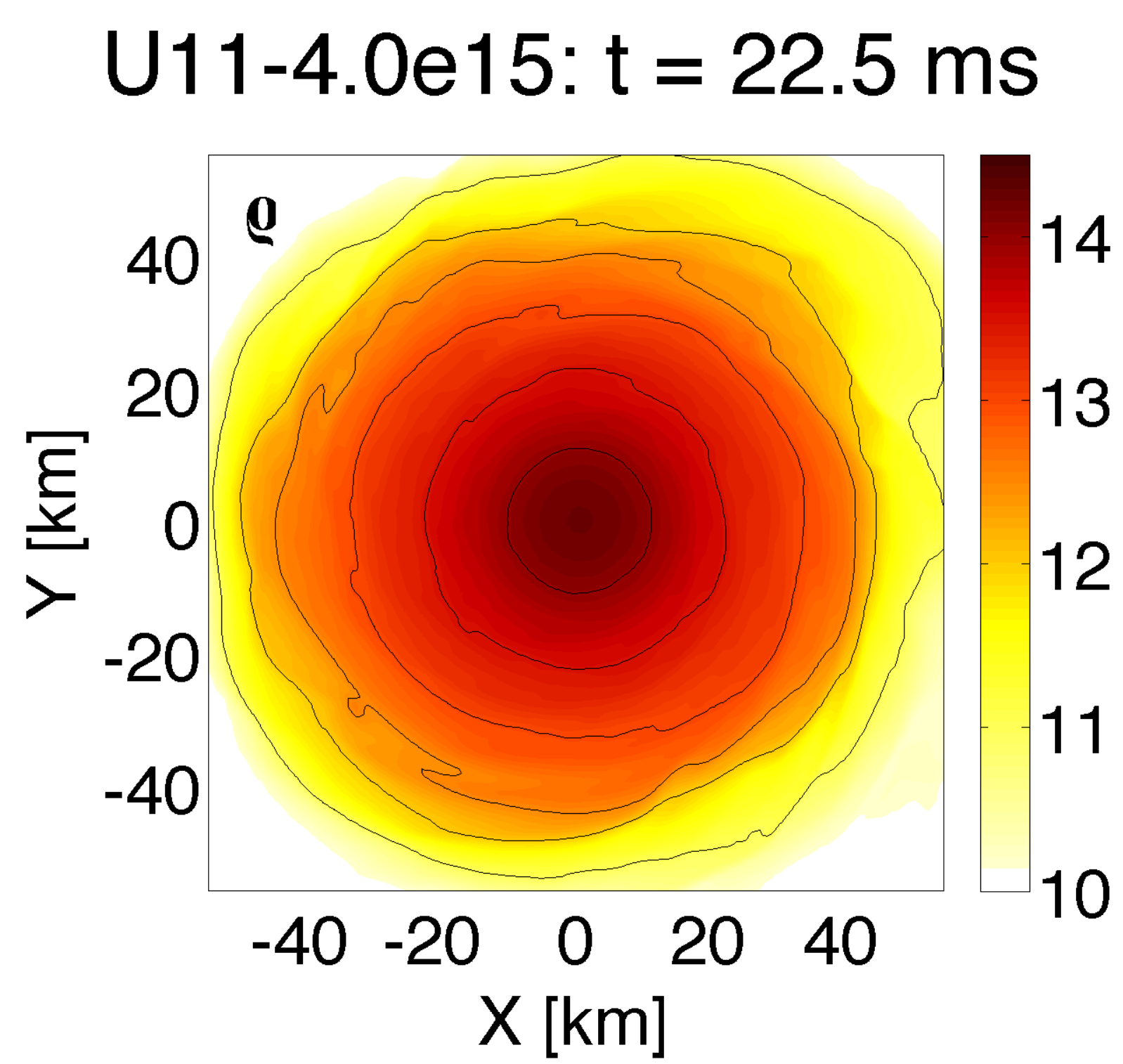}
\hspace{-0.1cm}
\includegraphics[width=0.24\textwidth]{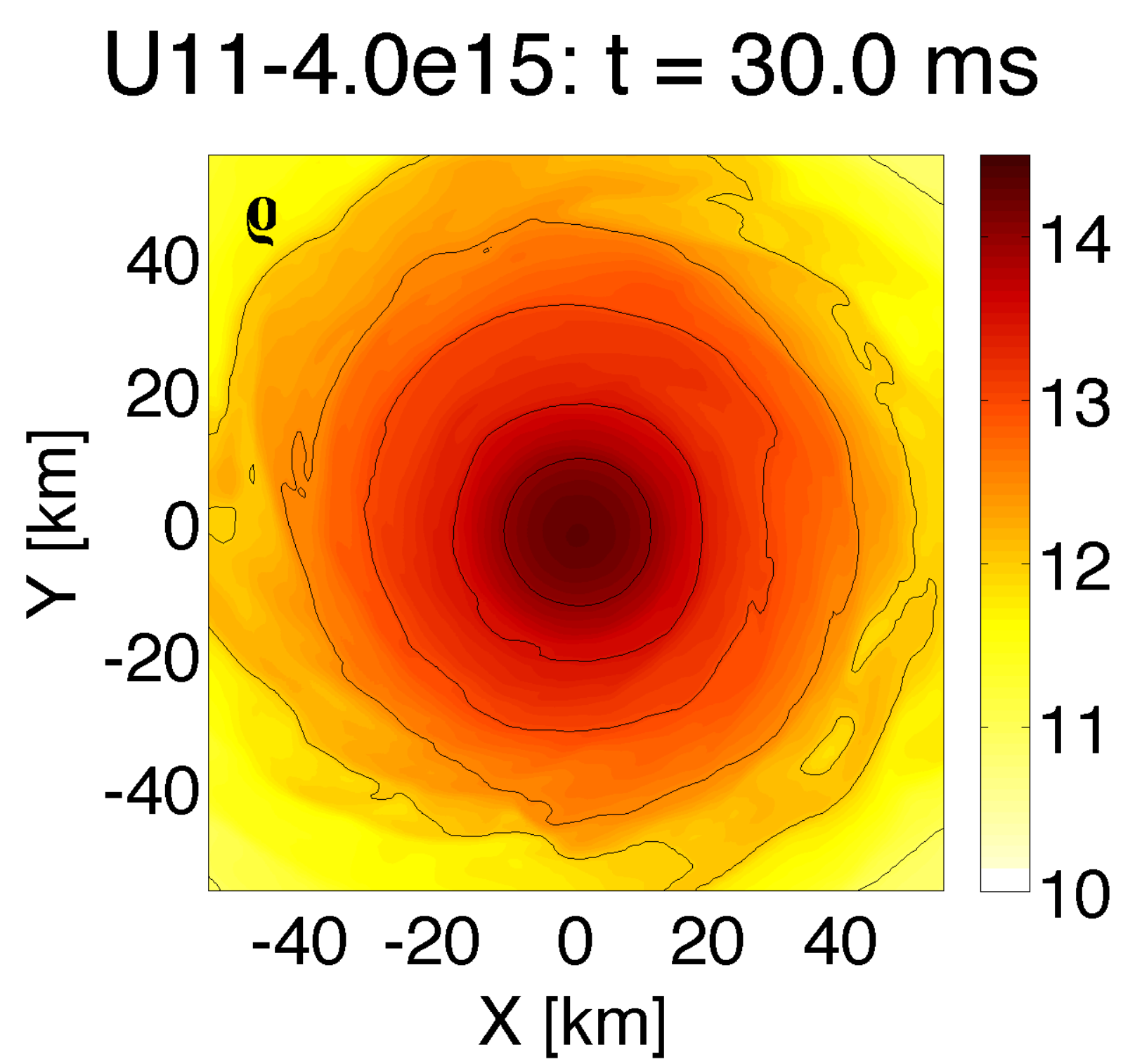}
}
\vspace{-0.1cm}\hbox{
\includegraphics[width=0.27\textwidth]{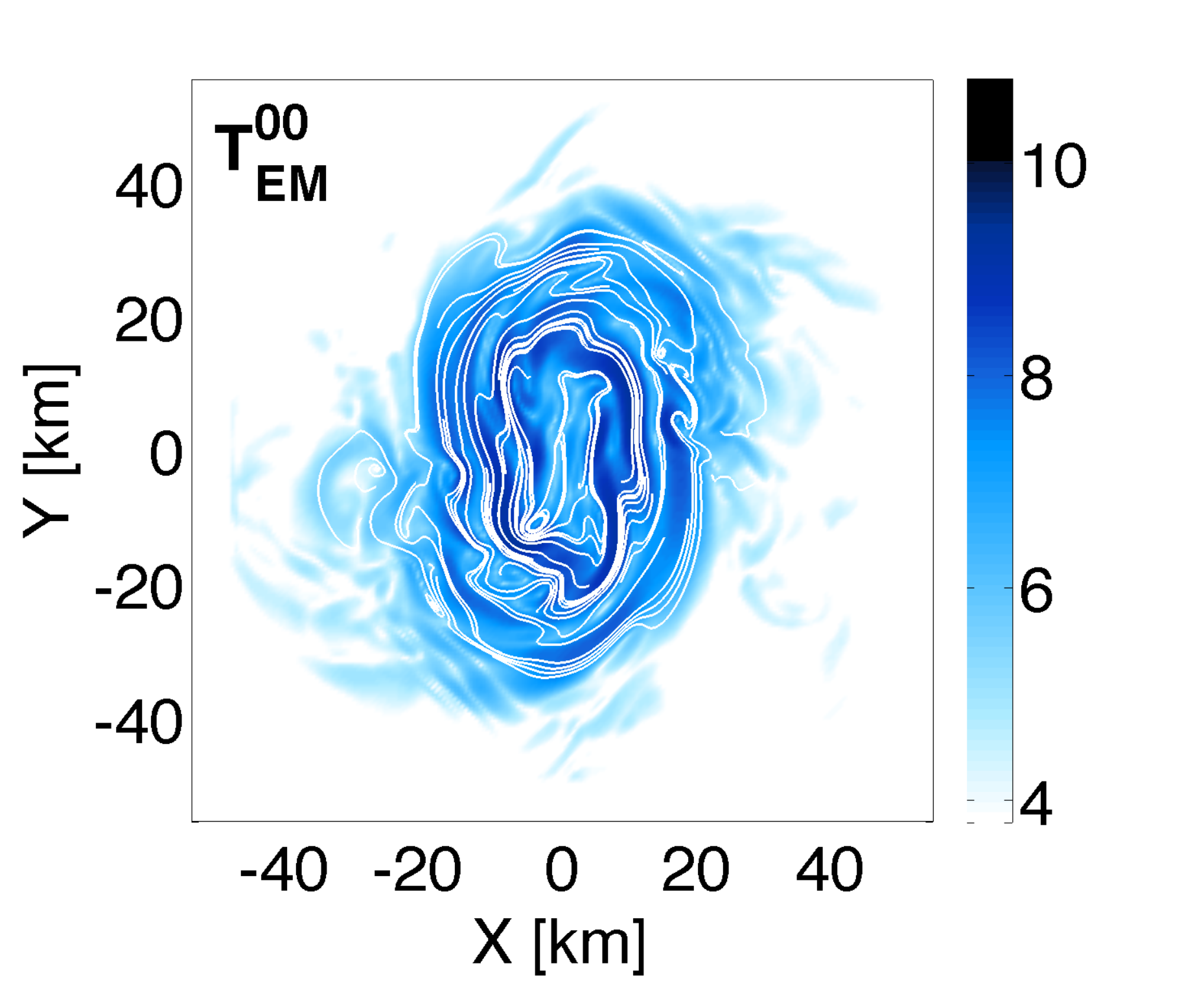}
\hspace{-0.6cm}
\includegraphics[width=0.27\textwidth]{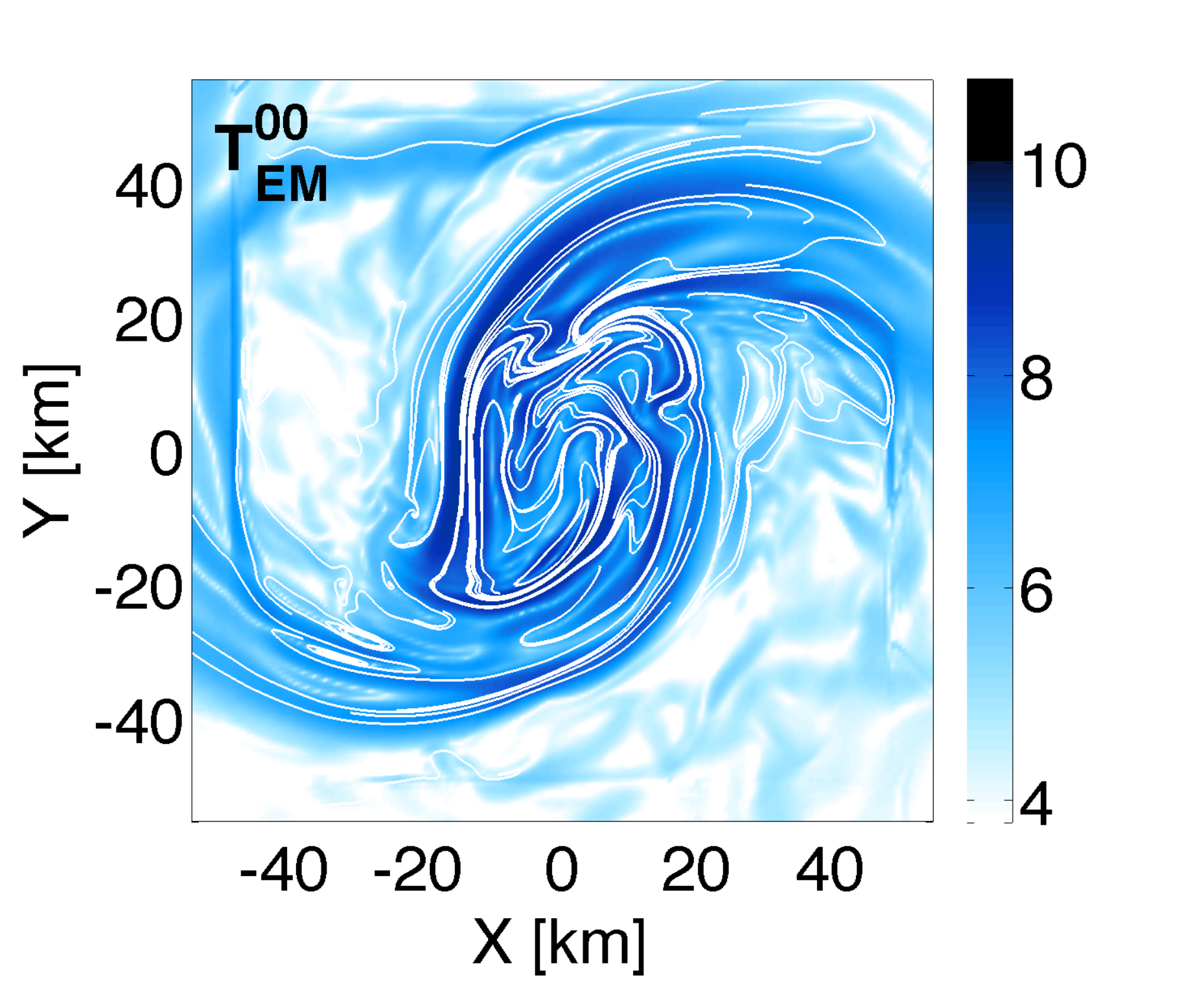}
\hspace{-0.6cm}
\includegraphics[width=0.27\textwidth]{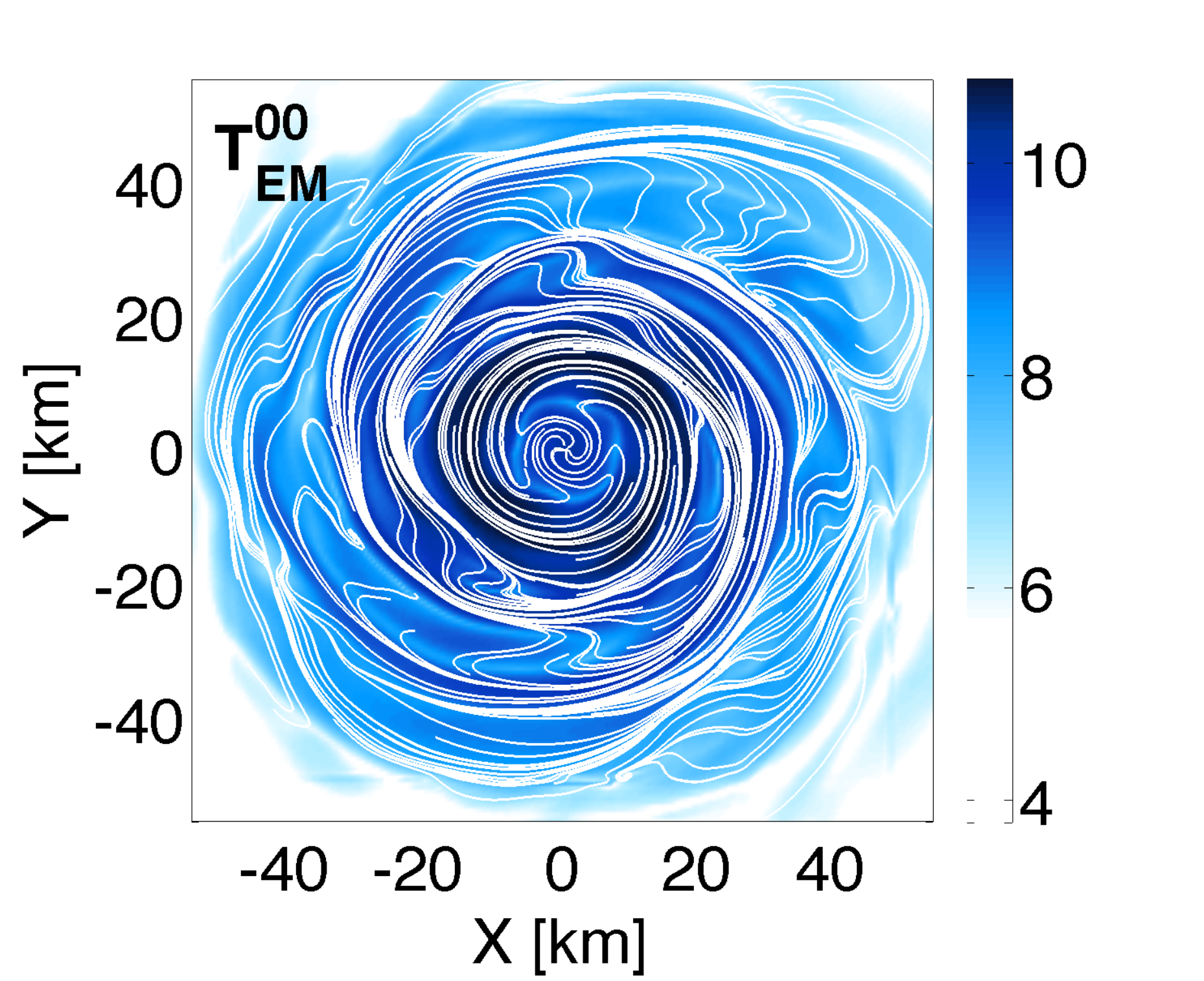}
\hspace{-0.6cm}
\includegraphics[width=0.27\textwidth]{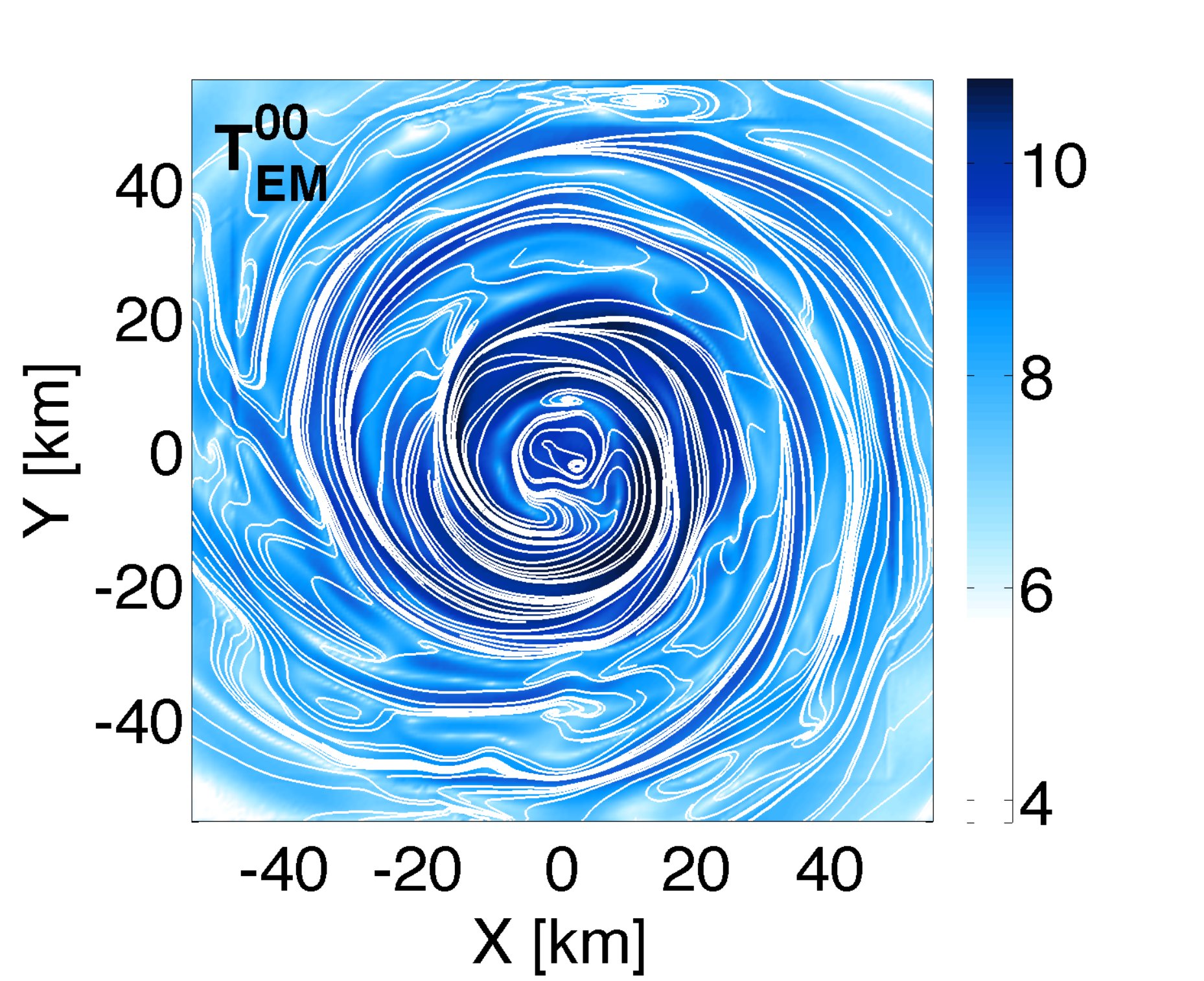}
}}
\end{center}
\vglue-0.9cm
\caption{(Top panels): snapshots of the rest-mass density $\rho$ on
  the $(x,\, y)$ plane at $t =$ 22.5 and 30.0 ms for model
  \texttt{U11} with two different seed magnetic field strengths:
  $B^{z}_\text{max}|_{t,z=0}=1.0 \times 10^{14}$ G for model
  \texttt{U11-1.0e14} (left panels) and $B^{z}_\text{max}|_{t,z=0}=4.0
  \times 10^{15}$ G for \texttt{U11-4.0e15} (right panels).  The color
  code is defined in terms of $\log_{10}(\rho)$ where $\rho$ is in cgs
  units (g/cm$^3$).  Additionally, isodensity contours are shown for
  $\rho = 10^{11}$, $10^{12}$, $5\times 10^{12}$, $10^{13}$, $5\times
  10^{13}$ and $10^{14}$ g/cm$^3$; (Bottom panels): snapshots of the local
  magnetic energy $T^{00}_{\text{EM}}$ on a horizontal plane at
  $z\simeq 1.5$ km for the same two models at the same steps during
  the evolution. Additionally, magnetic field lines are drawn in
  white.  The color code is defined in terms of
  $\log_{10}(T^{00}_{\text{EM}}/c^2)$ where $T^{00}_{\text{EM}}/c^2$
  is in cgs units (g/cm$^3$). }
\label{fig:RhoSnapshots}
\end{figure*}

\begin{figure*}[1.0\textwidth,t]
  \begin{center}
    \hspace{-0.35cm}
    \includegraphics[width=0.295\textwidth]{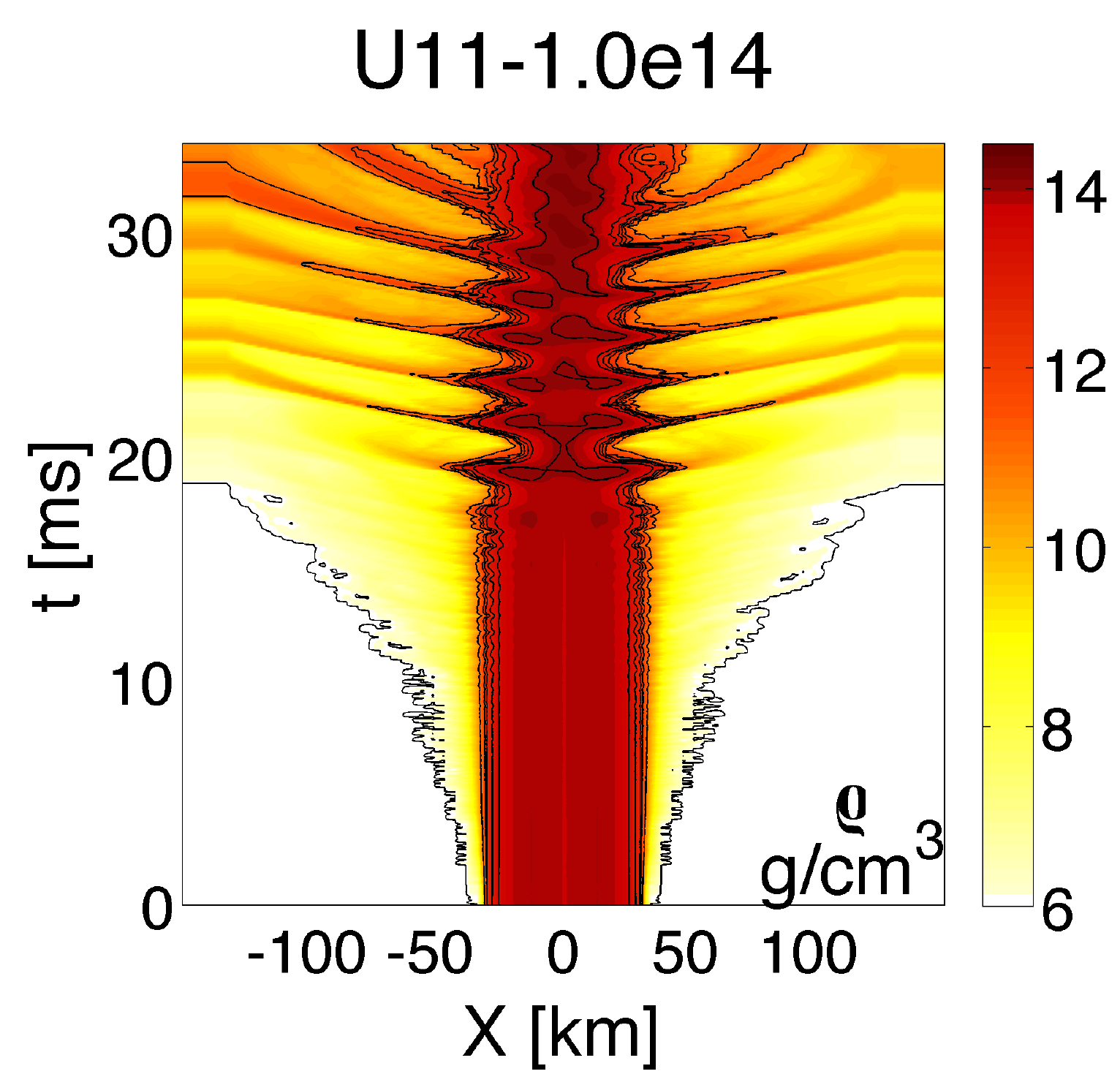}
    \hspace{0.28cm}
    \includegraphics[width=0.295\textwidth]{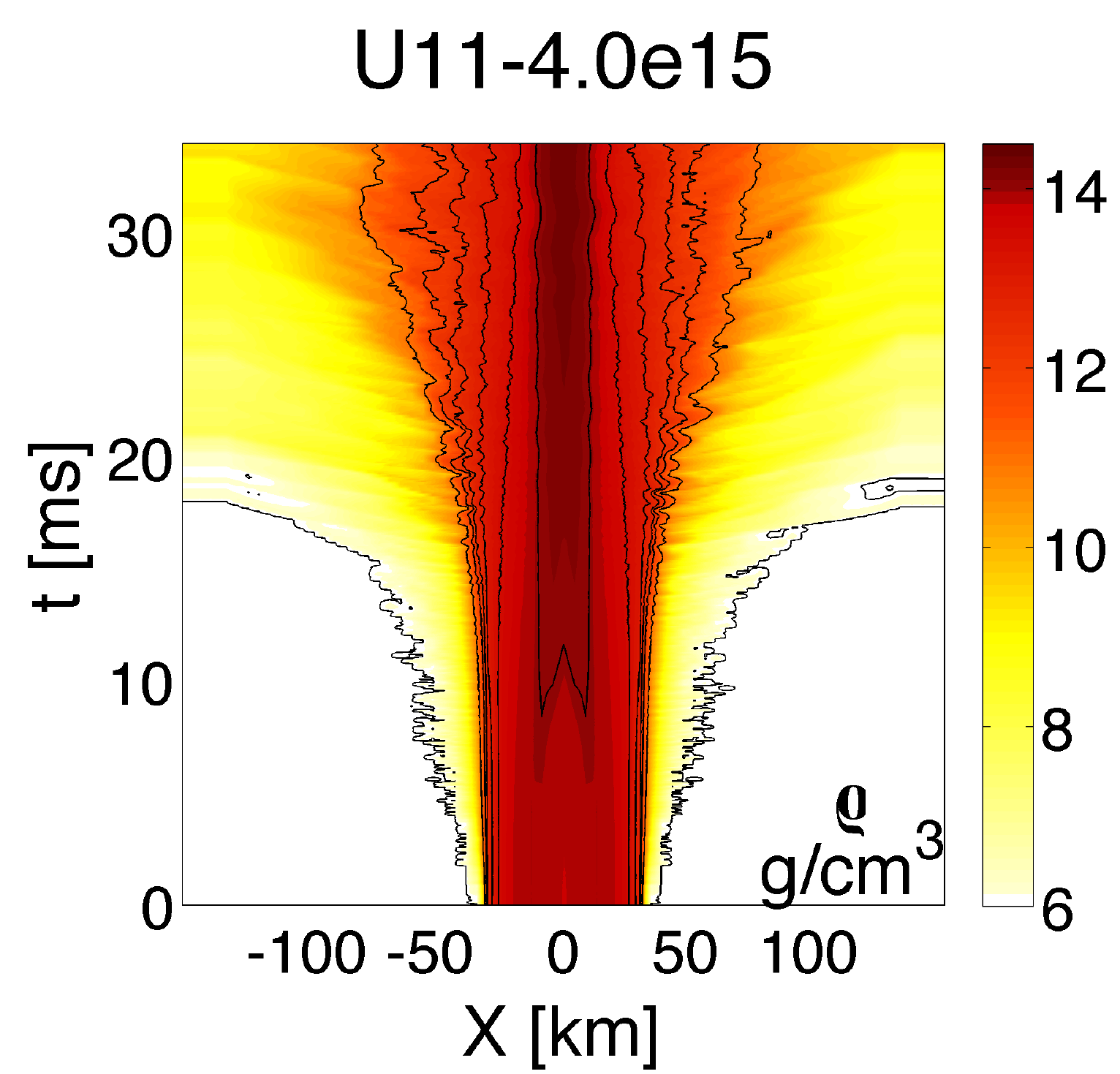}
    \hspace{0.28cm}
    \includegraphics[width=0.295\textwidth]{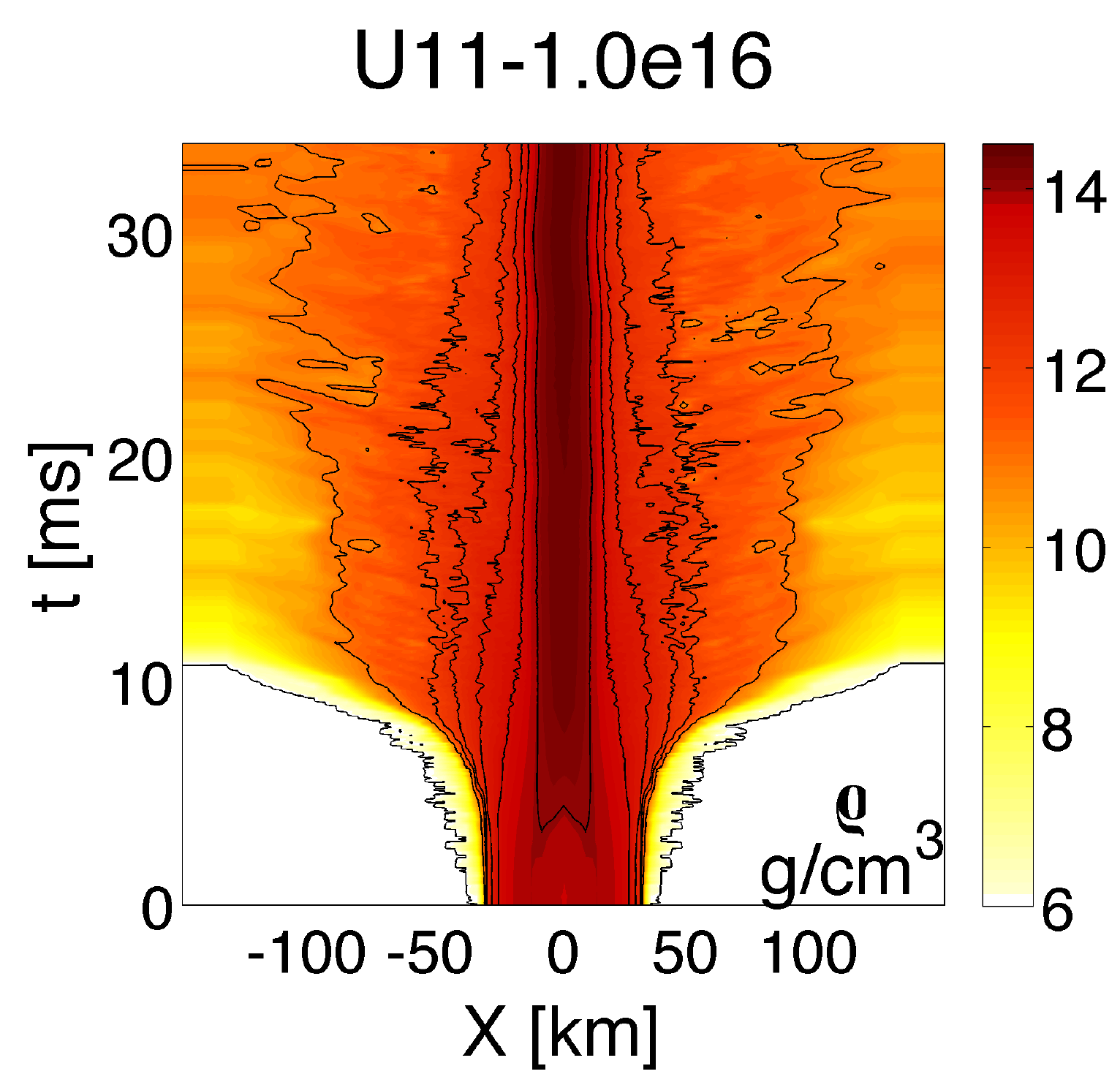}    
    \includegraphics[width=0.32\textwidth]{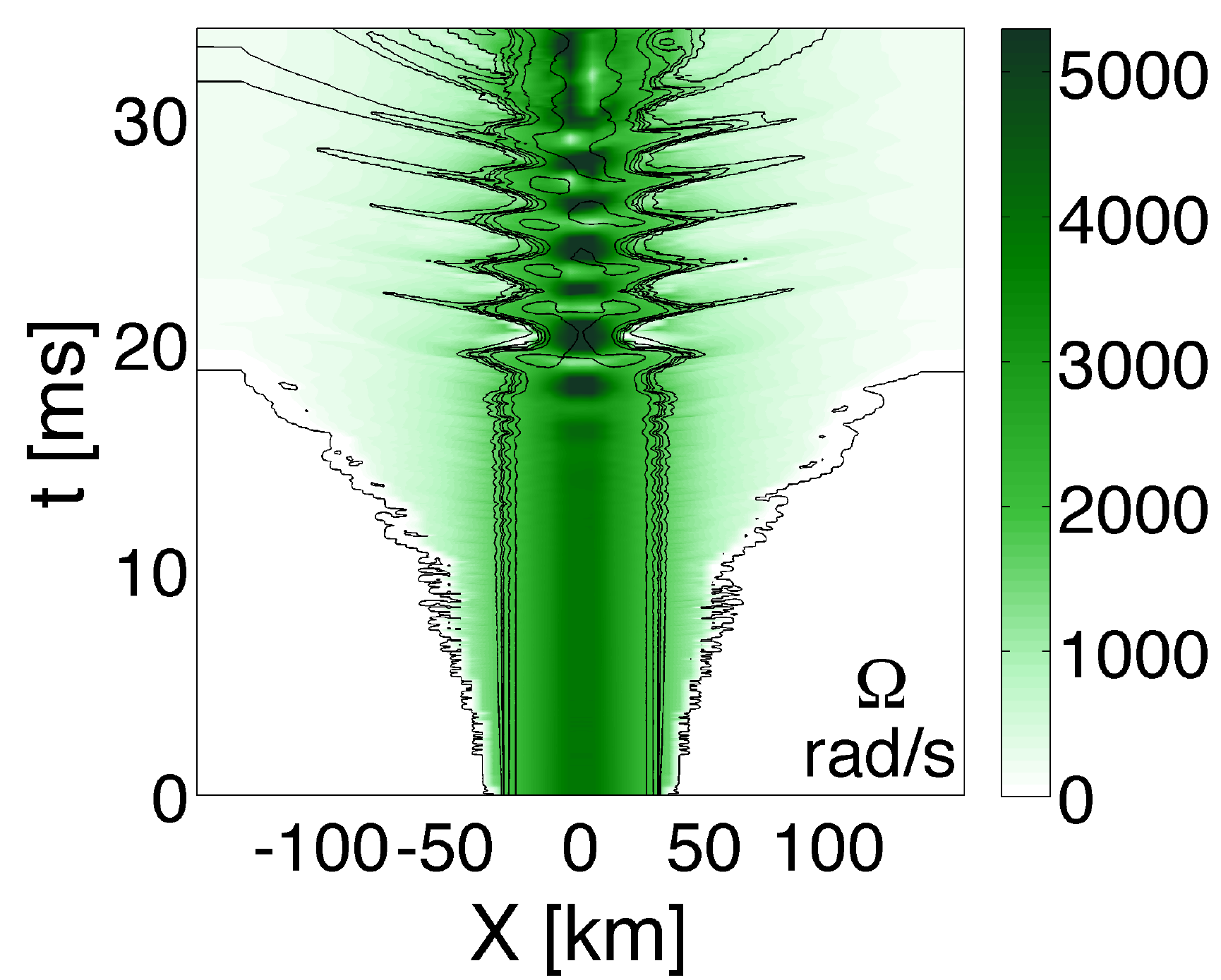}
    \includegraphics[width=0.32\textwidth]{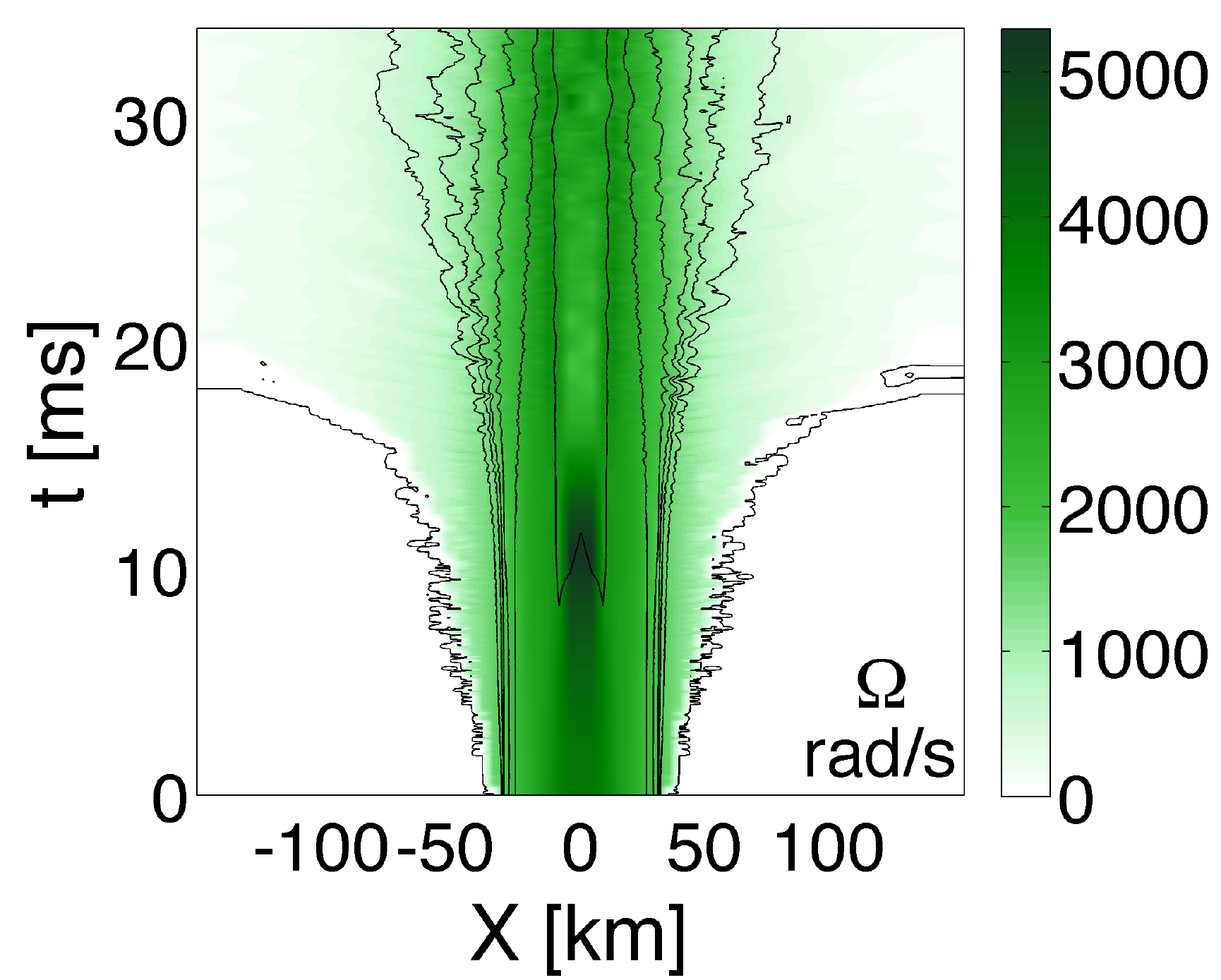}
    \includegraphics[width=0.32\textwidth]{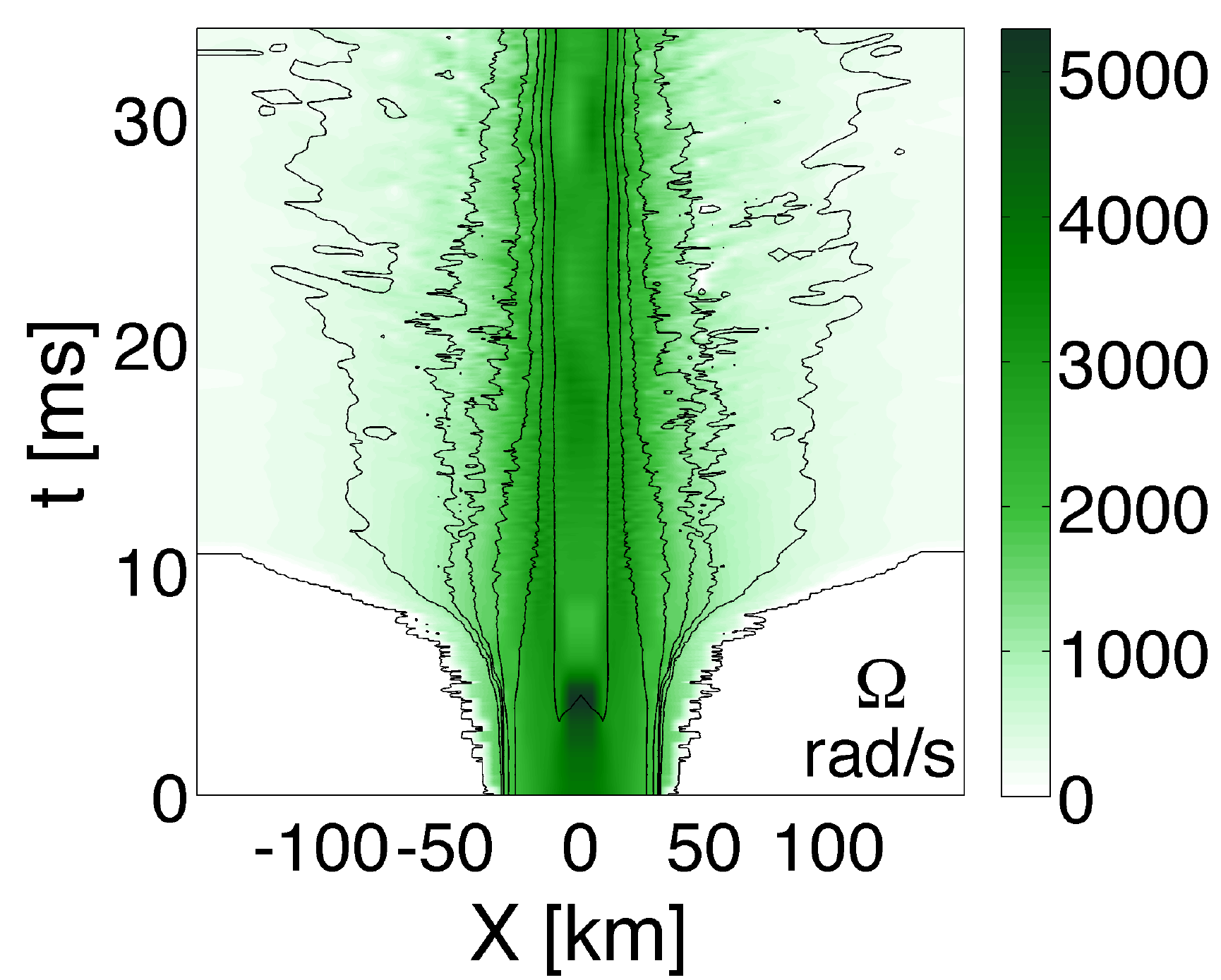}
  \end{center}
  \vglue-0.5cm
  \caption{Evolution of the rest-mass density $\rho$ (top row) and the
    angular velocity $\Omega$ (bottom row) along the $x$ axis for
    models \texttt{U11-1.0e14} (left panels), \texttt{U11-4.0e15}
    (central panels) and \texttt{U11-1.0e16} (right panels). The
    panels show the color-coded quantities embedded in a space-time
    diagram with the coordinate time $t$ on the vertical
    axis. Additionally, on top of all diagrams isodensity contours are
    shown for $\rho = 10^{6}, \; 10^{11}, \; 10^{12}, \; 5\times 10^{12}, \;
    10^{13}, \;5\times 10^{13}$ and $10^{14} \; \text{g/cm}^3$.}
  \label{fig:spacetime}
\end{figure*}

\section{Computational setup}
The simulations reported here have been performed using the
general-relativistic magnetohydrodynamics (GRMHD) code WhiskyMHD
described in~\cite{Bruno} and which is based on the {\tt Cactus} computational
toolkit. The gravitational fields are evolved using the BSSNOK formulation
with the same gauge conditions and parameters as in
\cite{Baiotti2007}, while the GRMHD equations are solved using a
high-resolution shock-capturing scheme based on the piecewise parabolic
(PPM) reconstruction and the Harten-Lax-van Leer-Einfeldt (HLLE)
approximate Riemann solver. For all the simulations discussed here we
have used a grid structure with four refinement levels, an outer boundary located at $L
\simeq 100 \; M_{\odot}$ ($ \simeq 150$ km) and finest resolution of 
$\Delta x = 0.375 \; M_{\odot}$ ($\simeq 0.5$ km). With this setting the outer
boundary is far enough to have all the dynamics happening inside the computational
domain, while the used resolution is already able to capture the dynamics of the magnetic
field (see \cite{Franci2013}).

\section{Results}

For each unstable model (\texttt{U3}, \texttt{U11} and \texttt{U13})
we have performed a number of simulations adding an initial purely
poloidal magnetic field around the typical value of the field
strengths expected for magnetars (that is of the order of $10^{15}$
G) and, indeed, in the range from $1.0 \times 10^{14}$ G to $1.0
\times 10^{16}$ G. Much stronger initial magnetic fields are quite
unlikely to be realistic, while weaker fields seems to have no effects
at all on the dynamics of the bar-mode instability. 

The dynamics of the bar-mode instability ($m = 2$ mode) is studied
following the evolution of the distorsion parameter $\eta
=(\eta_+^2+\eta_\times^2)^{1/2}$, which is defined in terms of the
matter quadrupole moment in the $xy$ directions through the 
quantities $\eta_\times=2 \,I^{xy}/(I^{xx}+I^{yy})$ and
$\eta_+=(I^{xx}-I^{yy})/(I^{xx}+I^{yy})$. In the central panel of
Fig.~\ref{fig:InitialModels} we show how the dynamics of $\eta$ is 
affected by different choices of the
initial magnetic field strength: in unstable
models, like model \texttt{U11-1.0e14}, this parameter shows an exponential growth
(whose growth time, $\tau_{bar}$, can be easly measured).

In particular, regarding the bar-mode instability, we draw the
conclusion that there are three distinct behaviors corresponding to
different seed magnetic fields.  

Indeed, we observe no effects at all
on the dynamics of the instability up to a certain magnetic field
strength (namely, $B^{z}_\text{max}|_{t,z=0} \lesssim 2.0 \times
10^{14}$ G for model \texttt{U3}, $4.0 \times 10^{14}$ G for model
\texttt{U11} and $8.0 \times 10^{14}$ G for model \texttt{U13}), since
both the growth time and the frequency of the m=2 matter mode do not
vary within the accuracy they can be computed with (see the left
panels of Fig.~\ref{fig:RhoSnapshots} and Fig.~\ref{fig:spacetime} and
the right panel of Fig.~\ref{fig:InitialModels}).  For higher field
strengths we observe a gradual slowdown in the development of the
instability, since the magnetic field lines are frozen into the fluid
and the differential rotation has to drag them together with the
matter implying that the instability growth time $\tau_{bar}$ is
higher for higher initial values of $\beta_{\text{mag}}$ (see the
right panel of Fig.~\ref{fig:InitialModels}). The models that are
still matter-unstable, even when a magnetic field is present, are
represented with squares in the left panel of
Fig.~\ref{fig:InitialModels} showing the parameter space
$(\beta,\;\beta_{\text{mag}})$.

Going to much higher fields we reach a threshold value above which the
instability is completely suppressed and the deformation no longer
develops. This threshold on the magnetic field still depends on the
initial model adopted, being higher than about $6.0 \times 10^{14}$ for model
\texttt{U3}, $2.0 \times 10^{15}$ for model \texttt{U11} and $2.4
\times 10^{15}$ for model \texttt{U13} (see Fig.~\ref{fig:InitialModels}). 
\begin{figure*}[1.0\textwidth,t!]
  \begin{center}
        \includegraphics[width=0.32\textwidth]{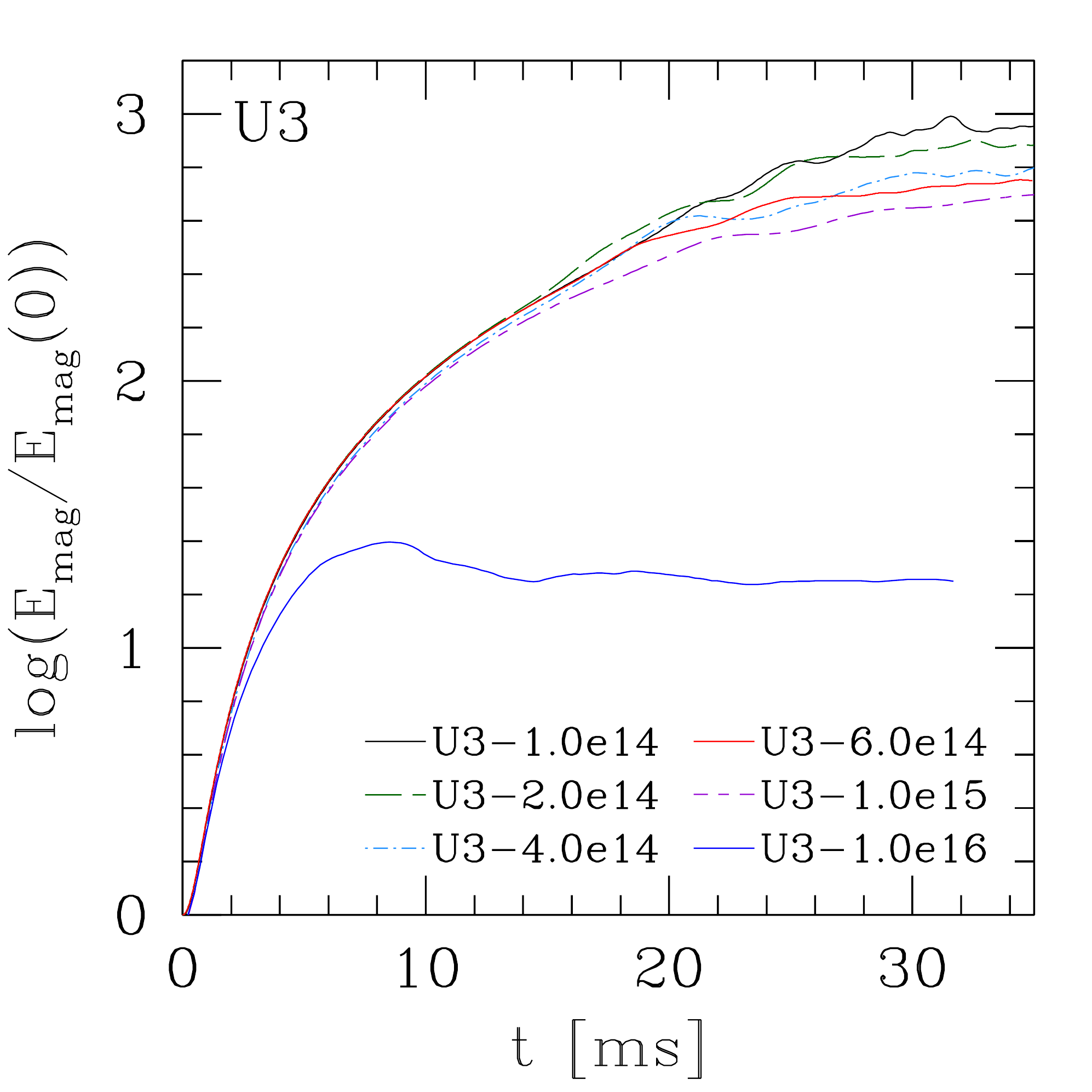}
        \includegraphics[width=0.32\textwidth]{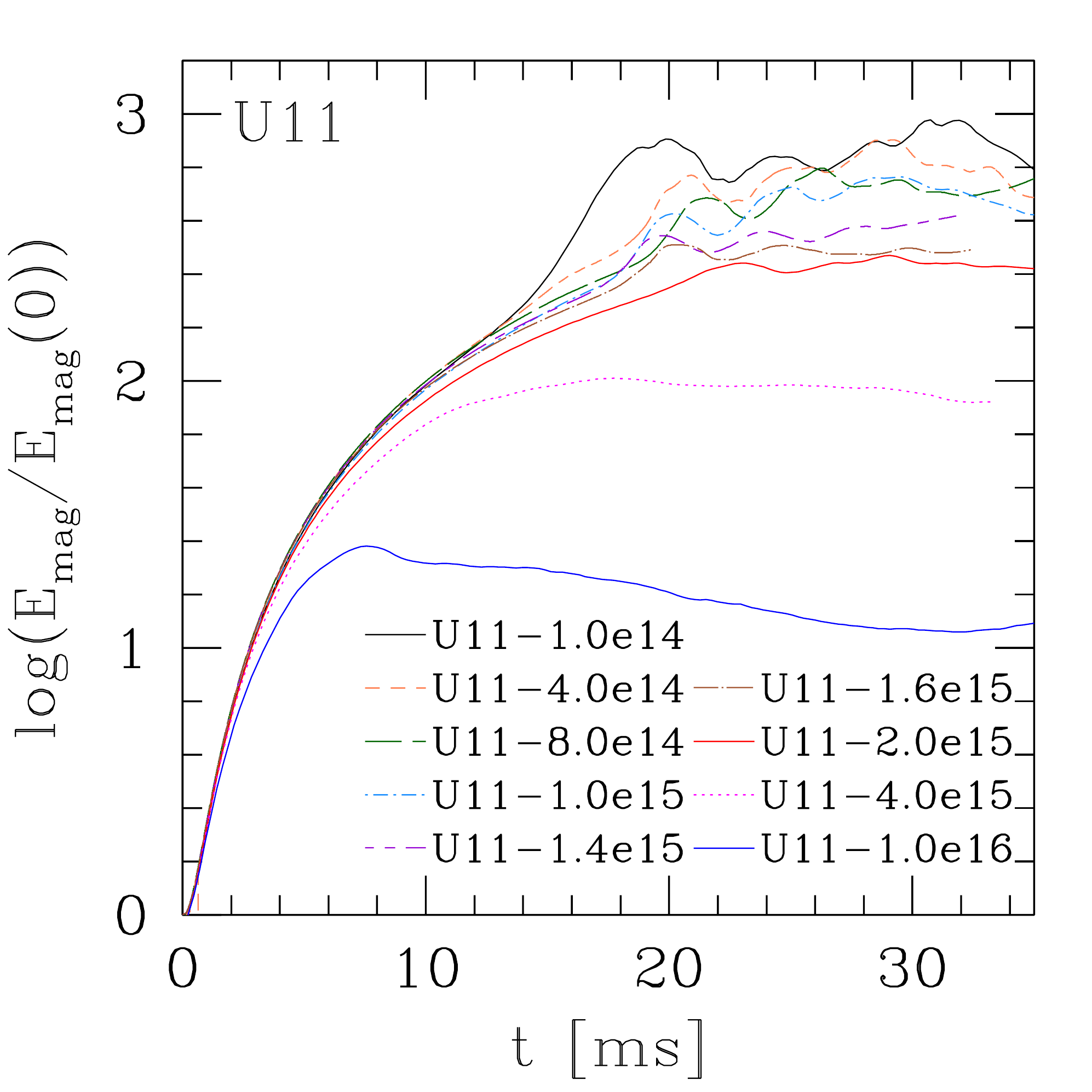}
        \includegraphics[width=0.32\textwidth]{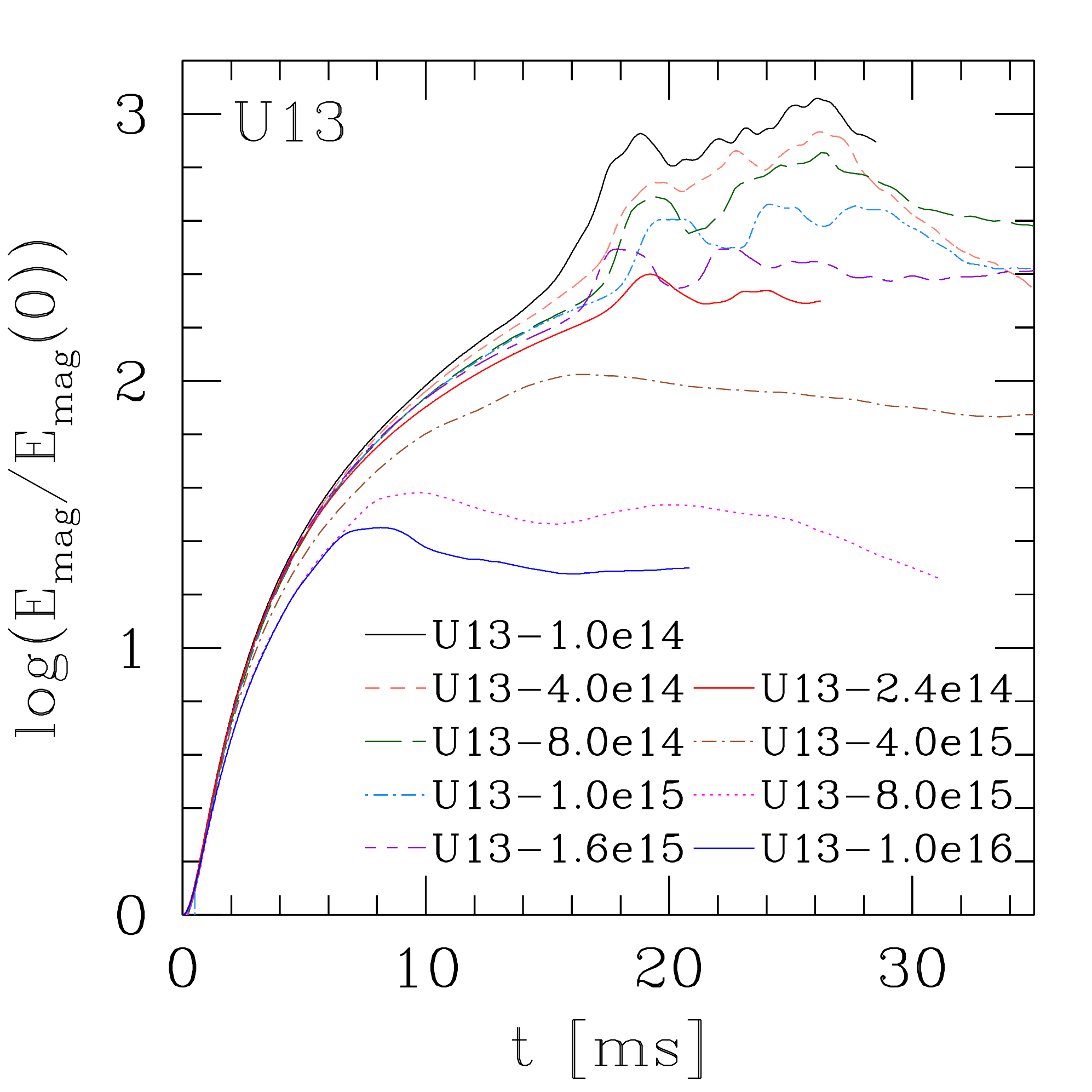}
  \end{center}
  \vglue-0.8cm
  \caption{Time evolution of the total magnetic energy
    $E_{\text{mag}}$ normalized to its initial value for the three
    models which are unstable in the non-magnetized case (\texttt{U3},
    \texttt{U11} and \texttt{U13}) corresponding to a wide range of
    seed magnetic fields from $B^{z}_\text{max}|_{t,z=0}=1.0 \times
    10^{14}$ G to $1.0 \times 10^{16}$ G. We use a black solid line
    to depict the less magnetized case, a blue solid line to indicate the most
    magnetized case and a red solid line for the last unstable model
    just before the suppression of the bar-mode instability due to the
    presence of the magnetic field.}
  \label{fig:Emag}
\end{figure*}

Besides, for very strong seed magnetic fields, we also observe changes
on the density and angular velocity profiles of the stellar models, so
in the end of the evolution we obtain a much more compact
configuration that is nearly uniformly rotating and surrounded by a very
low density envelope which is still differentially rotating (see the
right panels in Fig.~\ref{fig:RhoSnapshots} for model
\texttt{U11}). This is consistent with the expectation that magnetic
braking is transferring angular momentum from the core to the outer
layers. Models that are matter-stable are indicated by circles in Fig.
\ref{fig:InitialModels}, while the ones that also exhibit the
above-mentioned expansion of the outer layers due to the presence of
the magnetic field are represented with triangles.

Taking a closer look at the magnetic field evolution for all seed
magnetic fields, we observe that magnetic field lines wind due to
differential rotation, and hence a development and linear growth of a
toroidal component take place, with a consequent amplification of the
total magnetic energy of about two orders of magnitude or even more
(see Fig.~\ref{fig:Emag}). Moreover, we observe a further exponential
growth of both poloidal and toroidal components during the
matter-unstable phase, whose nature still needs to be investigated.

After having studied how the magnetic field affects the onset and
development of the bar-mode instability, we turned to investigate
whether there are effects on the dynamics of models which are
known to be stable in the non-magnetized case. Once again, this was
achieved by superimposing a purely poloidal seed magnetic field
allowing the magnetic field strength to take two values;
$B^{z}_\text{max}|_{t,z=0}=1.0 \times 10^{15}$ G and
$B^{z}_\text{max}|_{t,z=0}=1.0 \times 10^{16}$ G.

Regarding the distribution of matter we observe negligible effects when
$B^{z}_\text{max}|_{t,z=0}=1.0 \times 10^{15}$ G. Only for model
\texttt{S1}, which is very close to the threshold for the onset of the
dynamical instability (see the left panel of
Fig.~\ref{fig:InitialModels}), we observe some minor changes. More specifically, the
density profile turns from a toroidal initial configuration to a
nearly standard one with its maximum quite close to the $z$-axis. In
stable models with $B^{z}_\text{max}|_{t,z=0}=1.0 \times 10^{16}$ G,
instead, we observe the same behavior already described for highly
magnetized models, namely the outer layers expand forming an envelope
of much lower density around the initial stellar model.

\section{Conclusions}

We have analyzed how the presence of magnetic fields can affect the
development of the bar-mode $(m=2)$ instability in relativistic
differentially rotating stellar models with a seed poloidal magnetic field
in the range $10^{14}-10^{16}$ G. In order to do that we have performed 3D
magneto-hydrodynamical simulations coupled to the Einstein equations.

For all the models studied we have found, as expected, a sudden formation
and linear growth of a toroidal component of the magnetic field (in a
twisted configuration) that rapidly overcomes the original poloidal
one and  an amplification of the total magnetic energy to up to
almost three orders of magnitude. 

For relativistic stellar models that are
bar-mode unstable, we have observed almost no effects on the bar-mode
dynamics due to magnetic fields of the order of $10^{15}$ G or
below. For stronger magnetic fields the growth time of the instability
increases and the bar deformation appears to partially stall. Moreover,
magnetic fields stronger than about $10^{15}$ G are able to
completely suppress the instability and to change the density and
angular velocity profiles of the stellar models, eventually leading to
a final configuration where the star has almost uniform rotation
and is surrounded by a very low density envelope. 

he same
evolution is realized also for relativistic stellar models that are
stable against bar-mode deformations for seed magnetic fields of the order
of $10^{16}$ G. Moreover, we note that the presence of magnetic fields does
not alter the dynamics in stable models up to field strengths of the
order of about $10^{15}$ G.

The overall conclusion is that only magnetic fields of the order of
$10^{15}$ G or above seem to be able to suppress the purely
hydrodynamical instability and so it is quite unlikely that in
realistic astrophysical situations such a suppression due to the
presence of magnetic fields might occur. Indeed, the general picture
discussed in~\cite{Camarda2009} applies also in the non Newtonian regime.

This research has been possible thanks to the HPC resources of the
INFN ``Theophys'' cluster and to the PRACE allocation (6th-call)
``3dMagRoI'' on CINECA's Fermi supercomputer.

\smallskip

\end{document}